\documentclass[11pt]{article}
\usepackage{amsfonts, amssymb, graphicx, a4, epic, eepic, epsfig,
setspace, lscape}
\usepackage{color,soul,cancel}
\definecolor{Red}{cmyk}{0,1,1,0}
\definecolor{Blue}{cmyk}{1,1,0,0}
\definecolor{ForestGreen}{cmyk}{0.91,0,0.88,0.12}

\def\inst#1{$^{#1}$}
\def\ind#1{\1_{\{#1\}}}
\def\1{\rlap{\mbox{\small\rm 1}}\kern.15em 1}

\newtheorem{theorem}{Theorem}[section]
\newtheorem{lemma}[theorem]{Lemma}
\newtheorem{proposition}[theorem]{Proposition}

\newtheorem{definition}[theorem]{Definition}
\newtheorem{corollary}[theorem]{Corollary}
\newtheorem{remark}[theorem]{Remark}

\newcommand{\cX}{{\cal X}}

\def \s {{\sigma}}
\def \z {{\zeta}}
\def \g {{\gamma}}
\def \t {{\tau}}

\def \p {{\pi}}

\def\<{{\langle}}
\def\>{{\rangle}}

\newcommand{\be}[1]{\begin{equation}\label{#1}}
\newcommand{\ee}{\end{equation}}

\newcommand{\bl}[1]{\begin{lemma}\label{#1}}
\newcommand{\el}{\end{lemma}}

\newcommand{\br}[1]{\begin{remark}\label{#1}}
\newcommand{\er}{\end{remark}}

\newcommand{\bt}[1]{\begin{theorem}\label{#1}}
\newcommand{\et}{\end{theorem}}

\newcommand{\bd}[1]{\begin{definition}\label{#1}}
\newcommand{\ed}{\end{definition}}

\newcommand{\bcl}[1]{\begin{claim}\label{#1}}
\newcommand{\ecl}{\end{claim}}

\newcommand{\bp}[1]{\begin{proposition}\label{#1}}
\newcommand{\ep}{\end{proposition}}

\newcommand{\bc}[1]{\begin{corollary}\label{#1}}
\newcommand{\ec}{\end{corollary}}

\newcommand{\bi}{\begin{itemize}}
\newcommand{\ei}{\end{itemize}}

\newcommand{\ben}{\begin{enumerate}}
\newcommand{\een}{\end{enumerate}}

\begin {document}

%%%%%%%%%%%%%%%%%%%%%%%%%%%%%%%%%%%%%%%%%%%%%%%

%%%%%%%%%%%%%%%%%%%%%%%%%%%%%%%%%%%%%%%%%%%%%%%%%

\title{Effects of boundary conditions on irreversible dynamics}

\author{
Aldo Procacci\inst{1}\and
Benedetto Scoppola\inst{2} \and
Elisabetta Scoppola\inst{3}}

%\date{}

\maketitle

\begin{center}
{\footnotesize
\vspace{0.3cm} \inst{1}  Departamento de Matem\'atica,
Universidade Federal de Minas Gerais \\
Av. Ant\'onio Carlos, 6627 -
30161-970  Belo Horizonte,  Brazil\\
\texttt{aldo@mat.ufmg.br}\\

\vspace{0.3cm}\inst{2}  Dipartimento di Matematica, Universit\`a di Roma
``Tor Vergata''\\
Via della Ricerca Scientifica - 00133 Roma, Italy\\
\texttt{scoppola@mat.uniroma2.it}\\

\vspace{0.3cm} \inst{3} Dipartimento di Matematica e Fisica, Universit\`a
Roma Tre\\
Largo San Murialdo, 1 - 00146 Roma, Italy\\
\texttt{scoppola@mat.uniroma3.it}\\ }

\end{center}

\vskip3.truecm

\begin{abstract}
\noindent
We present a simple one-dimensional Ising-type spin system on which
we define a completely asymmetric Markovian single spin-flip dynamics.
We study the system  at a very low, yet non-zero,  temperature and we
show that for empty
boundary conditions the Gibbs measure is stationary for such dynamics,
while introducing in a single site a $+$ condition the stationary measure
changes drastically, with macroscopical effects. We achieve this result defining an
absolutely convergent series expansion of the
stationary measure around the zero temperature system. Interesting combinatorial identities are
involved in the proofs.
\end{abstract}

\eject

%\tableofcontents
%%%%%%%%%%%%%SECT 1%%%%%%%%%%%%%%%%%%%%%%%%%
\section{Introduction}
\label{Intro}
\vspace{0.3cm}

In this paper we discuss a very simple one-dimensional spin system in order to point out the crucial
effect of boundary conditions on the invariant measure of irreversible dynamics.

\noindent
Irreversible dynamics turn out to be
 a challenging problem since they are the main ingredient in the study of non-equilibrium statistical mechanics.
Indeed many interesting  physical systems can not be described in terms
of equilibrium: for instance  non-Hamiltonian evolutions,  systems with external non-conservative forces,
or systems with thermostats or reservoirs.
Such systems exhibit non zero currents of matter or energy flowing in an {\it irreversible} way.
For this kind of problems it is necessary
to consider non-equilibrium statistical mechanics. Actually we can say that
the description of non-equilibrium systems represents one of the ``grand challenges" in statistical mechanics.

\noindent
In this frame the main point is to describe {\it Non-Equilibrium Stationary States } (NESS), ``in understanding the properties of states which are in stationary nonequilibrium: thus establishing a clear separation between properties of evolution towards stationarity (or equilibrium) and properties of the stationary states themselves: a distinction which until the 1970's was rather blurred."
as mentioned in the beautiful book by Gallavotti \cite{G}.

\noindent
Irreversible dynamics play in this context a crucial role.
The invariant measures of  irreversible dynamics are
stationary states but they describe non zero currents of probability, and hence they are NESS.
A famous example is given by the  TASEP model, in which particles hop only to the right, entering from a left reservoir with a given  rate  and leaving the
system from the site $L$ with another rate.

\noindent
In the context of Markovian dynamics, given any two states $i$ and $j$
in some configuration space $\mathcal X$,  the irreversibility is
defined by transition probabilities $P(i,j)$ violating the detailed balance condition
$$\p(j)P(j,i)=\p(i)P(i,j)\qquad \forall i,j\in \cX$$
This means that there are non zero probability currents. Indeed given a pair of states $i,j\in \cX$
 define
the probability current (or flow of probability) from $j$ to $i$ at time $t$ the asymmetric function
on $\cX\times\cX$:
$$
K_t(j,i)=P^t(j)P(j,i)-P^t(i)P(i,j)
$$
where $P^t(\cdot)$ represents the probability of the state $\cdot$ at time $t$.

\noindent
The continuity equation for $P^t(i)$,  gives
$$
P^{t+1}(i)-P^t(i)=\sum_jP^t(j)P(j,i)-P^t(i)\sum_jP(i,j)=$$
$$
=\sum_{j\not=i}\Big(P^t(j)P(j,i)-P^t(i)P(i,j)\Big)=\sum_{j\not=i}K_t(j,i)=-({\rm div} \, K_t)(i)
$$

Stationarity implies
 \be{inv_meas}
0=\sum_{j\not=i}\Big(\p(j)P(j,i)-\p(i)P(i,j)\Big)=\sum_{j\not=i
}K(j,i) \qquad \forall i \ee being $K(j,i)=\p(j)P(j,i)-\p(i)P(i,j)$,
the stationary probability current (or stationary flow of
probability) from $j$ to $i$, a divergence free flow. This flow $K$
is proportional to the antisymmetric part of the conductance
associated to the chain and it is also considered for instance in
\cite{GL}. Actually the presence of  currents can be used to detect
irreversible dynamics without using the invariant measure. This is
done by the {Kolmogorov criterion for reversibility} {\cite{Ko}}:
the Markov dynamics with transition probabilities $P(i,j)$ is
reversible if and only if for any loop of states: $i_o, i_1, i_2,
..., i_n, i_o$ we have
$$
P(i_0,i_1)P(i_1,i_2)....P(i_n,i_0)=P(i_0,i_n)....P(i_2,i_1)P(i_1,i_0).
$$
This means  that the dynamics is irreversible if there is a loop with a stationary current.
As noted in the rich review by Chou, Mallick and Zia,  \cite{CMZ}, the presence of stationary current loops suggests
to associate magnetostatics  to irreversible dynamics as electrostatics is associated to
reversible dynamics.
\bigskip

\noindent
Beside their crucial role  in the understanding of
non-equilibrium statistical mechanics,  irreversible dynamics have been frequently considered in the literature
in order
to speed up  simulations. Indeed in some case rigorous control
of mixing time of irreversible dynamics has been obtained. See for instance \cite{dss2}.

\bigskip
\noindent
{Several problems arise when considering   irreversible dynamics.}
{Indeed some tools frequently used in the study of convergence to equilibrium  are strongly related to reversibility}
as
spectral representation or  the  potential theoretical approach.
Recently some progress has been done to extend some of these
tools to non reversible dynamics.  See for instance the extension of the
Dirichlet principle to non reversible Markov chains obtained  in \cite{GL}.

\noindent
In this paper we want to stress the main difficulty related to irreversibility:  while
detailed balance is a crucial tool to control the invariant measure of reversible dynamics,
in the irreversible case the
control of the invariant measure can be quite complicated,
and in particular it is difficult to study its sensitivity
to boundary conditions. Very recent results have been obtained in this direction in
\cite{gab} where irreversible dynamics are constructed with a given Gibbsian stationary measure
by exploiting cyclic decomposition of divergence free flows.

\noindent
In some case it is possible to verify that the equation for the invariant measure (\ref{inv_meas})
is satisfied by a suitable Gibbs measure, as proved below in the (easy) case of empty boundary
conditions. This is also the case of 2-dimensional Ising model with asymmetric interaction
discussed in \cite{pss} with periodic boundary condition.
In general, due to the presence of probability currents,
the verification of equation (\ref{inv_meas}) typically involves non local argument
and so the invariant measure strongly depends on boundary conditions.

\bigskip

\noindent
We consider a one dimensional spin system on the discrete interval $[1,L]\equiv\{1, 2, ..., L\}$ with a single-spin-flip
Markovian dynamics $\{X_t\}_{t\in\mathbb{N}}$,
defined on $\cX:=\{-1,1\}^L$ by the following transition
probabilities
\be{def-P}
P(\s,\s^{(i)})=\frac{1}{L} e^{-2J(\s_i\s_{i-1}+1)}
\ee
where $\s^{(i)} $ is the configuration obtained from $\s$ flipping the spin
in the site $i\in\{1,2,...,L\}$.
This means that at each time a site $i$ is chosen uniformly at random in $\{1,2,...,L\}$ and the spin is flipped in
this site with probability one if it is opposite to its left neighbour, $\s_{i-1}$, or with probability $e^{-4J}$ if
it is parallel to $\s_{i-1}$.
%We will denote by $\P_\s(.)$ the probability measure on the space of paths of  $X_t$
%starting at $X_0=\s$.
We will consider two different boundary conditions:
\bi
\item[-] the empty boundary condition corresponding to $\s_0=0$;
\item[-] the + boundary condition corresponding to $\s_0=+1$.
\ei
The chain is irreducible and aperiodic so that in both cases there exists a unique invariant measure.
Our goal is to compare the invariant measures of the Markov chains corresponding to
these two different boundary conditions in a very low temperature regime, i.e., when the parameter $J$ is sufficiently large
w.r.t. $L$.

\noindent
We shall prove that while in the case of empty boundary conditions the stationary distribution
is the Gibbs measure, in the case of $+$ boundary condition the stationary measure
changes drastically. Due to the particular low-temperature regime we are able to write
the stationary distribution as an absolutely convergent expansion in $e^{-4J}$.
This expansion is easily controlled in this case, but it could be a general tool in order
to control the invariant measure at a very low temperature in more general contexts.
We control completely the first order of such expansion, and we show that it has
several interesting features. In particular, the presence of probability currents implies
that the boundary conditions do not have the effect of a conditioning, as in the
case of the Gibbs measure. The
boundary conditions actually modify the stationary distribution
and the effect of their presence decay very slowly in the distance $i$ from the boundary,
namely as $\frac{1}{\sqrt{i}}$. Moreover, the presence of boundary conditions makes
the probabilities of interval of minus spins dependent on their length, producing
macroscopical effects on the magnetization.

\noindent
The paper is organized as follows: in section 2 we define the models, comparing them with the
usual reversible Glauber Dynamics for the 1d Ising Model, and we state the main results of the paper.
Section 3 is devoted to the control of the expansion of the invariant measure in terms of the
quantity $e^{-4J}$. Section 4 contains the proof related to the characterization of the
first order term of the invariant measure.
Some conclusion remarks and future perspectives are discussed in section 5.

%%%%%%
\section{Models and results}
\label{0bc}
As mentioned in the introduction, our model is
defined via an irreversible markovian dynamics on a one dimensional discrete spin chain with
states $\s\in{\cal X}=\{-1,+1\}^{\{1,2,\dots, L\}}$.
 We consider two different boundary conditions, namely the free boundary conditions, having $\s_0=\s_{L+1}=0$, and the $+$ boundary condition
 $\s_0=\s_{L+1}=+1$. The dynamics is defined by the following transition matrix
\be{irre}
P^I(\sigma,\tau)=\cases{{1\over L}e^{-2J(\sigma_i\sigma_{i-1}+1)}& if $\tau=\sigma^{(i)}$ \cr
1-{1\over L}\sum_i e^{-2J(\sigma_i\sigma_{i-1}+1)}& if $\tau=\sigma$\cr
0&otherwise\cr}
\ee
where $\s^{(i)}$ is the configuration obtained from $\s$ by flipping the spin in the site $i$.
This dynamics is irreversible, but in the case of free boundary condition it is
easy to find its stationary measure.
Indeed, consider the Gibbs measure

\be{def-Gm}
\p^G(\s) = \frac{e^{-H(\s)}}{Z^G}, \qquad Z^G=\sum_{\s\in\cX}e^{-H(\s)}
\ee
where $H(\s)$ is the usual Ising Hamiltonian with free boundary conditions.
\be{def-H}
H(\s)=-J\sum_{i=2}^L \s_i\s_{i-1}
\ee
Let us show that $\p^G(\s)$ is the unique stationary measure of dynamics (\ref{irre}). The dynamics is
clearly irreducible and aperiodic, and hence the stationary measure exists and it is unique.

%The transition probabilities between adjacent states
%$\s, \s^{(i)}$ may be rewritten in the following way
%
%
%\be{irr2}
%P^I_{\s,\s^{(i)}}=\cases{
%{1\over L}e^{-4J}& for $\sigma_i\sigma_{i-1}=1,\ i\ge2$\cr
%{1\over L}e^{-2J}& for $i=1$\cr
%{1\over L}& otherwise\cr}
%\ee
%

\noindent
Moreover it is immediate to verify the following equalities:
\be{1.1}
\p^G(\s^{(i)})=\p^G(\s) e^{-2J(\s_i\s_{i-1}+\s_i\s_{i+1})}
\ee
\be{def-Pirr}
P^I(\s^{(i)},\s)=\frac{1}{L} e^{2J(\s_i\s_{i-1}-1)}.
\ee
To prove that $\p^G$ is the invariant measure of the process
satisfying
\be{BC1}
\sum_{\t\in\cX}\p^G(\t)P^I(\t,\s)=\p^G(\s)
\ee
it is sufficient to verify
the following  condition, obtained by (\ref{BC1}) by canceling the diagonal terms in both
sides of the equality, which is equivalent to equation (\ref{inv_meas}):
\be{BC}
\sum_{i=1}^L\p^G(\s^{(i)})P^I(\s^{(i)},\s)=\p^G(\s)\sum_{i=1}^LP^I(\s,\s^{(i)})
\ee
Equation (\ref{BC}) immediately follows from (\ref{1.1}) and (\ref{def-Pirr}) since, by the empty b.c we have
$$
\sum_{i=1}^L e^{-2J(\s_i\s_{i+1})}=\sum_{i=1}^{L-1} e^{-2J(\s_i\s_{i+1})}=\sum_{i=2}^L e^{-2J(\s_i\s_{i-1})}.
$$
It is a standard task to define a reversible markovian dynamics having the same  stationary measure, i.e. the
well known {\it Glauber dynamics}, given by the following
transition probability matrix
\be{glau}
P^R(\s,\t)=\cases{\frac{1}{L}e^{-[H(\s^{(i)})-H(\s)]_+}& if  $\t=\s^{(i)}$\cr
1-\sum_i\frac{1}{L}e^{-[H(\s^{(i)})-H(\s)]_+}& if  $\t=\s$\cr
0& otherwise
}
\ee
where $[\cdot]_+$ means the positive part.

\noindent
%To help the comparison with the transition probabilities defined in (\ref{irr2})
%note that the transition probabilities between adjacent states
%$\s, \s^{(i)}$ may be rewritten in the following way
%
%\be{glau2}
%P^R_{\s,\s^{(i)}}=\cases{\frac{1}{L}e^{-4J}& for $\s_i=\s_{i-1}=\s_{i+1},\ i\ge2$\cr
%\frac{1}{L}e^{-2J}& for $\s_1=\s_{2},\ i=1$\cr
%\frac{1}{L}& otherwise\cr
%}
%\ee
%

\noindent
For both dynamics the one-dimensional stationary measure $\p^G(\s)$ is well known.
We have

$$
\p^G(\s)=\frac {e^{-2J\ell(\s)}}{2\Big(1+e^{-2J} \Big)^{L-1}}
$$
where $\ell(\s)$ is the number of pair $\{i,i+1\}$ such that $\s_i\s_{i+1}=-1$ (i.e. $\ell(\s)$ is the
total length of the Peierls contours).

\noindent
We conclude this short discussion of the  empty boundary conditions
by checking the irreversibility of this dynamics, i.e., the presence of non zero probability currents. Indeed,
for example, for $i>1$ and $m>1$ such that $i+m<L$,  let us consider the configuration
$\s$ with $\s_j=-1$ for $j=i, i+1,\dots, i+m-1$ and $\s_j=+1$ elsewhere and observe that $\p^G(\s)=\p^G(\s^{(i)})$ while
$P(\s,\s^{(i)})={1\over L}$ and $P(\s^{(i)},\s)=  {e^{ -4J}\over L}$. Therefore
$$
\p^G(\s)P(\s,\s^{(i)})- \p^G(\s^{(i)})P(\s^{(i)},\s)= {1- e^{ -4J}\over L} \p^G(\s)>0
$$

\noindent
In order to control the invariant measure in the case of plus boundary conditions,
we introduce a particular regime, defined as follows.
\vglue.3cm
\noindent
{\bf Definition}.

\noindent
We say that
the one-dimensional discrete spin chain on $[1,L]$ with
states $\s\!\in\!\{-1,\!+1\}^{\{1,\dots, L\}}$ subjected to the irreversible dynamics (\ref{irre}) or to
the Glauber  dynamics (\ref{glau}) is
in the
{\it chilled regime of parameter $c>0$ if
$$J=c\log L$$}

\noindent
Note that the Gibbs measure $\p^G$ for $c$ large enough
is concentrated on the configurations $\s=\boxplus$ ($\s_i=1\ \forall i$) and
$\s=\boxminus$ ($\s_i=-1\ \forall i$), while for the other configurations
$\s$ we get
$$
\p^G(\s)\sim\frac{1}{2}{e^{-2J\ell(\s)}}
$$

\noindent
The chilled condition defined above mimics a phase transition, in the sense
that the volume dependent low temperature (high $J$) defined by  $e^{-2J}L\ll 1$ forces
the system in a non zero (in particular, very close to $\pm1$) magnetization.
It is very easy, yet quite interesting, to study
the mixing time of the two dynamics defined above, which is proportional to
the expected value of the {\it tunneling time}, namely,
the time needed to pass from the configuration $\boxplus$
to the configuration $\boxminus$.

\noindent
It is not difficult to identify in the reversible case the typical path of the tunneling.
By chilled condition
$e^{-2J}L\ll 1$, a spin flip on the boundary occurs
after a time of the order of $L e^{2J}$and a spin flip inside a region of spins
having all the same sign occurs after a time of the order of $e^{4J}$.
Both times are much longer
than $L$.
On the other side the interface between two regions with opposite spins
may move in a time of order $L$, with equal probability on the right and on the left.
Hence the typical path of the tunneling is a spin flip on one of the two boundaries
followed
by a random walk of the boundary between the $+$ and the $-$ regions which eventually reaches
the other boundary.  The latter
event happens with probability $1/L^2$, giving in the end a tunneling
time of the order of $L^3 e^{2J}$.
In the irreversible dynamics the spin in the site $1$ is flipped
after a time of the order $L e^{2J}$. The boundary between the $+$ and the $-$ regions, then,
typically moves only on the right, and this
ensures that the tunneling time is of the order of $L e^{2J}$, and
hence shorter, polynomially in $L$, than the reversible case.

\vskip.4cm
\noindent
In what follows we will consider the case of  $+$boundary conditions, namely $\s_0=\s_{L+1}=1$. With the
reversible Glauber dynamics (\ref{glau})   the invariant measure  with plus boundary conditions is just Gibbs measure $\pi^G$ conditioned
to $\s_0=\s_{L+1}=1$. If we consider now the irreversible dynamics (\ref{irre}) we will see ahead that
its invariant measure   changes dramatically with respect to the free boundary conditions case.

\noindent
For notational simplicity in the computation we will also use the notation
$P_{\s\t}\equiv P(\s,\t)$ and $\pi_\s\equiv \pi(\s)$.

%%%%%%%%%%%%%%%%%%%%%%%
\subsection{Results}
\label{pert}

\noindent
Before stating our results concerning this particular regime we need to introduce the main technical tool which
consists in writing the invariant measure of the irreversible dynamics with $+$boundary conditions in the chilled regime
in terms of  a series  in $e^{-4J}$.
We will omit for simplicity hereafter the
suffix $I$, standing for irreversibility.

\noindent
Denoting with $\ell(\s)$ the number of antiparallel pairs of spins
for each  configuration $\s$ and recalling that $\s_0=1$,
we can write the transition probability matrix in the following form
\be{trans+}
P(\sigma,\tau)=\cases{{1\over L}& if $\tau=\sigma^{(i)}$ and $\sigma_i\sigma_{i-1}=-1$\cr
{e^{-4J}\over L}& if $\tau=\sigma^{(i)}$ and $\sigma_i\sigma_{i-1}=1$\cr
1-{\ell(\s)\over L}-\left(1-{\ell(\s)\over L}\right)e^{-4J}& if $\tau=\sigma$\cr
0&otherwise\cr}
\ee

We can define the dynamics above for zero temperature ($J\rightarrow\infty$)

$$P^{(0)}(\sigma,\tau)=\cases{{1\over L}& if $\tau=\sigma^{(i)}$ and $\sigma_i\sigma_{i-1}=-1$\cr
1-{\ell(\s)\over L}& if $\tau=\sigma$\cr
0&otherwise\cr}$$
obtaining
\be{perturbP}
P(\sigma,\tau)=P^{(0)}(\sigma,\tau)+e^{-4J}\Delta P(\sigma,\tau)
\ee
where
\be{delta}\Delta P(\sigma,\tau)=\cases{{1\over L}& if $\tau=\sigma^{(i)}$ and $\sigma_i\sigma_{i-1}=1$\cr
-1+{\ell(\s)\over L}& if $\tau=\sigma$\cr
0&otherwise\cr}
\ee

\noindent
The state $\s$ corresponding to  $\ell(\s)=0$, i.e., $\sigma=\boxplus\  (\sigma_i=+1\ \forall\ i)$, is clearly
absorbent for the zero temperature dynamics. Hence
$$\pi^{(0)}(\sigma)=\cases{{1}& if $\sigma=\boxplus$\cr
0&otherwise\cr}$$

\noindent
We can use now the following formula for the perturbations on Markov chains:
\be{perturb1}
\pi(\sigma)=\sum_{k=0}^\infty e^{-4Jk}\pi^{(k)}(\sigma)
\ee
where
\be{perturb2}\pi^{(k)}(\sigma)=\sum_\tau\pi^{(0)}(\tau)D^k(\tau,\sigma)\qquad D=\sum_{j=0}^\infty \Delta P (P^{(0)})^j\ee

Again for notational simplicity we will write
$P^{(0) j}\equiv (P^{(0)})^j$.
Note that by its definition
\be{terminek}
\pi^{(k)}(\sigma)=0\quad \forall\ \s:\;\ell(\s)>2k
\ee

\noindent
Formulas (\ref{perturb1}) and (\ref{perturb2}) may be easily proved
in general. Indeed,
let $\pi^{(0)}_i$ the stationary measure of an ergodic Markov chain $P^{(0)}_{ij}$.
Consider the chain $P_{ij}=P^{(0)}_{ij}+\varepsilon \Delta P_{ij}$.
Denote with
$\pi_i$ the stationary measure of the chain $P_{ij}$. By ergodic theorem we have

$$\pi_i=\lim_{N\rightarrow\infty}\sum_j \pi^{(0)}_j (P_{ij})^N=
 \lim_{N\rightarrow\infty}\sum_j \pi^{(0)}_j (P^{(0)}_{ij}+\varepsilon \Delta P_{ij})^N$$
Then defining
$$D_{ij}=\sum_{l\ge 0}\sum_k\Delta P_{ik}(P^{(0) l})_{kj}$$
we have that
$$\pi_i=\sum_k\pi_i^{(k)}\varepsilon^k$$
with
$$\pi_i^{(k)}=\sum_l\pi^{(0)}_l(D^k)_{li}$$
A similar expansion  is used for instance in \cite{CLST} for the blockage problem.
\bigskip

\noindent
We define the expansion of the stationary measure up to the first order as

\be{pile1}\pi^{(\le 1)}=\pi^{(0)}+e^{-4J}\pi^{(1)}\ee
Note that $\pi^{(\le 1)}$ is a probability measure.

\noindent
We can now state our main results.
The first  is an immediate consequence of the convergence of the perturbative expansion (\ref{perturb1}).
Let
$$
d_{TV}(\pi,\pi^{(\le 1)})=\sum_\sigma\left|\pi(\s)- \pi^{(\le 1)}(\s)\right|
$$
be the total variation distance between the measure $\pi$ and its first order approximation $\pi^{(\le 1)}$.  Then the following theorem holds.

\bt{t1}
In chilled regime of parameter $c=\frac{1}{2}+\g$, with $\g>0$,
we have that
\be{th1} d_{TV}(\pi,\pi^{(\le 1)})\le \frac{const}{L^{8\g}}
\ee
\et
\noindent

\vskip.2cm
\noindent
Theorem \ref{t1}  shows that it is meaningful, in the
chilled regime with $\g>1/2$, to compute the first order in $e^{-4J}$ of the stationary
measure, since it will be the leading one.

\noindent As it is clear from the perturbative approach, by
(\ref{terminek}), up to first order the only configurations admitted
are the ones with at most one connected interval of sites having
$\s_i=-1$, while all the rest of the configuration has $\s_i=+1$.

\noindent
Let  $i\in {1,\dots,L-1}$ and $m\in{1,\dots, L-i}$ and let us denote  $(i;m)$ the state having
$$\s_k=\cases{+1&for $1\le k< i$\cr
-1&for $i\le k< i+m$\cr +1&for $i+m\le k\le L$\cr}
$$
In other words
$\sigma_i\sigma_{i-1}=-1$, $\sigma_{i+m}\sigma_{i+m-1}=-1$, $\sigma_k\sigma_{k-1}=1$ $\forall k\ne i,i+m$.  That is, the state $(i;m)$
is a single interval of $m$ spins equal to -1 starting at $i$,

\noindent
Let us denote $(i)$ the state having
$$\s_k=\cases{+1&for $1\le k< i$\cr
-1&for $i\le k\le L$\cr}
$$
In other words $(i)=(i;L+1-i)$, i.e.,
$\sigma_i\sigma_{i-1}=-1$, $\sigma_k\sigma_{k-1}=1$ $\forall k\ne i$.

\noindent

\bt{t2}
For any fixed $m>0$ and $i$ large we have
\be
{th'2ij}\pi^{(\le 1)}_{(i;m)}= e^{-4J}\left(1-\frac{C_m}{{\sqrt i}}+o\Big(\frac{1}{{\sqrt i}}\Big)\right)
\ee
where $C_m$ is a constant depending on $m$.
For every $i, m$  we have
\be{th2ij}\pi^{(\le 1)}_{(i;m)}\le { 4}\,e^{-4J} \;e^{-\frac{(m)^2}{2(i+m)}}~m
\ee
Moreover for every $i$
\be{th2i}\pi^{(\le 1)}_{(i)}=\sum_{l=1}^i\pi^{(\le 1)}_{(l;L-l)}
\ee
\et
\noindent
{\bf Remark}.
Note that by (\ref{th'2ij}) we get $\pi^{(\le 1)}_{(i;m)}\to e^{-4J}$ as $i\to\infty$, so that very far from the boundary
condition the stationary distribution at the first order in $e^{-4J}$ is equal to  the Gibbs one, giving the same weight
to every interval of minus spins independently of its length and its position. This convergence to
the Gibbs measure, however, it is very slow,  and it does not occur on a well defined scale.
Moreover the  exponential decay  with  the length $m$ of the interval of
$-$ spins given by (\ref{th2ij})  produces  macroscopic effects, as the following theorem shows.

\bt{t3}
The average value of $m(\s):=\sum_{i=1}^L\ind{\{\s_i=-1\}}$ with respect to the Gibbs measure, $\pi^G$, and with respect
to the irreversible measure up to the first order, $\pi^{(\le 1)}$, are such that
\be{th3}
\lim_{L\rightarrow\infty}\frac{\pi^{(\le 1)}(m)}{\pi^G(m)}\le\frac{1}{4}
\ee
\et

\section{Proof of Theorem \ref{t1}}

\noindent
By (\ref{perturb2}) we have

$$d_{TV}(\pi,\pi^{(\le 1)})=\sum_\sigma\left| \sum_{k=2}^\infty e^{-4Jk}\pi^{(k)}_{\sigma} \right|\le
\sum_{k=2}^\infty e^{-4Jk}\sum_\sigma\left|\pi^{(k)}_{\sigma} \right|$$
For $J=c\log L$ the condition $c=\frac{1}{2}+\g$ implies $e^{-4Jk}=L^{-(2+4\g)k}$
and then it is enough to prove that
\be{est}
\sum_\sigma\left|\pi^{(k)}_{\sigma} \right|\le(CL^2)^{k}\ee
Since
$$\sum_\sigma\left|\pi^{(k)}_{\sigma} \right|=\sum_\sigma
\left| \sum_{m=0}^\infty\sum_{\tau,\sigma'}\pi^{(k-1)}_{\tau}\Delta P_{\tau\sigma'}\left( P^{(0)m}\right)_{\sigma'\sigma}\right|$$
we have that (\ref{est}) is recursively proved if we are able to prove that

\be{stima}\sup_\tau\sum_{\sigma}
\left| \sum_{m=0}^\infty\sum_{\sigma'}\Delta P_{\tau\sigma'}\left( P^{(0)m}\right)_{\sigma'\sigma}\right|\le CL^2
\ee
Note first that
\be{normd}
\sum_{\sigma'}\Delta P_{\tau\sigma'}=0\ee
for all $\tau$. Then define
the matrix $\Pi^{(0)}$, having all the rows equal to the stationary measure $\pi^{(0)}$, and hence
having on the column related to the configuration $\s=\boxplus$, say on the first column,
all the entries equal to 1,
while all the other entries are zero.
Observe that, due to (\ref{normd}), we have
\be{normt}
\sum_{\sigma'}\Delta P_{\tau\sigma'}\Pi^{(0)}_{\s',\t}=0
\ee
for all $\s$ and $\t$.
Finally define
\be{erre}
R_m=P^{(0)m}-\Pi^{(0)}
\ee
Due to (\ref{normt}) we have that
\be{sumerre}
\sum_{\sigma'}\Delta P_{\tau\sigma'}\left(P^{(0)m}\right)_{\sigma'\sigma}=
\sum_{\sigma'}\Delta P_{\tau\sigma'}\left(R_m\right)_{\sigma'\sigma}
\ee

\noindent
 Now using (\ref{sumerre}) we split the sum over $m$ in two:
 $$\sum_{\sigma}
 \left|\sum_{m=0}^\infty \sum_{\sigma'}\Delta P_{\tau\sigma'}\left( P^{(0)m}\right)_{\sigma'\sigma}\right|\le$$
\be{agg}\le \sum_{\sigma}
 \left| \sum_{\sigma'}\Delta P_{\tau\sigma'}\left(\sum_{m=0}^{2L^2} P^{(0)m}\right)_{\sigma'\sigma}\right|+
 \sum_{\sigma}
 \left| \sum_{\sigma'}\Delta P_{\tau\sigma'}\left(\sum_{m=2L^2+1}^\infty R_m\right)_{\sigma'\sigma}\right|
 \ee
The first term is estimated as follows
 $$\sum_{\sigma}
 \left| \sum_{\sigma'}\Delta P_{\tau\sigma'}\left(\sum_{m=0}^{2L^2} P^{(0)m}\right)_{\sigma'\sigma}\right|\le\sum_{\sigma,\sigma'}
 \left| \Delta P_{\tau\sigma'}\left(\sum_{m=0}^{2L^2} P^{(0)m}\right)_{\sigma'\sigma}\right|\le$$
$$\le\sum_{\sigma'}
 \left| \Delta P_{\tau\sigma'}\right|\sum_{m=0}^{2L^2} \sum_\sigma
 \left(P^{(0)m}\right)_{\sigma'\sigma}$$
 The sum on $\sigma$ is 1 for each addend of the sum on
 $m$, and then
 $$\sum_{\sigma}
 \left|\sum_{\sigma'} \Delta P_{\tau\sigma'}\left(\sum_{m=0}^{2L^2} P^{(0)m}\right)_{\sigma'\sigma}\right|\le
 \sum_{\sigma'}
 \left| \Delta P_{\tau\sigma'}\right|2L^2
 $$
 Since, due to the definition of $\Delta P_{\tau\sigma'}$, we have
 \be{sommaD}
 \sum_{\sigma'}
 \left| \Delta P_{\tau\sigma'}\right|=2(1-{\ell(\t)\over L})\le 2
 \ee
 we obtain the following estimate
 \be{sommapiccola}\sum_{\sigma}
 \left| \sum_{\sigma'}\Delta P_{\tau\sigma'}\left(\sum_{m=0}^{2L^2} P^{(0)m}\right)_{\sigma'\sigma}\right|\le 4L^2
 \ee

 \noindent
 Now we are left with the estimate of the second term in (\ref{agg}):
 $$\sum_{\sigma}
 \left| \sum_{\sigma'}\Delta P_{\tau\sigma'}\left(\sum_{m=2L^2+1}^\infty R_m\right)_{\sigma'\sigma}\right|\le \sum_{\sigma, \s'}
 \left| \Delta P_{\tau\sigma'}\left(\sum_{m=2L^2+1}^\infty R_m\right)_{\sigma'\sigma}\right|$$

 \noindent
 Let us first of all consider the entries of the matrix $R_m$.
 Calling $T_\boxplus(\s')$ the hitting time to the state $\boxplus$ starting from the state
 $\s'$ we have that,
 being $\boxplus$ an absorbent state,
 $$(R_m)_{\s',\boxplus}= P^{(0) m}_{\s',\boxplus}-1=-P(T_\boxplus(\s')>m)$$
 For the same reason
 $$\sum_{\s\ne \boxplus}(R_m)_{\s',\s}=P(T_\boxplus(\s')>m)$$
 and therefore
 $$\sum_{\s}|(R_m)_{\s',\s}|=2P(T_\boxplus(\s')>m)$$
 Hence

 $$\sum_{\sigma, \s'}
 \left| \Delta P_{\tau\sigma'}\left(\sum_{m=2L^2+1}^\infty R_m\right)_{\sigma'\sigma}\right|
 \le 2\sum_{\sigma'}\left|  \Delta P_{\tau\sigma'}\sum_{m=2L^2+1}^\infty P(T_\boxplus(\s')>m)\right|\le
 $$
 $$
 \le 2\left(
 \sup_{\s'}\sum_{m=2L^2+1}^\infty P(T_\boxplus(\s')>m)\right)
 \sum_{\sigma'}\left| \Delta P_{\tau\sigma'}\right|\le
4 \sup_{\s'}\sum_{m=2L^2+1}^\infty P(T_\boxplus(\s')>m)$$
where in the last line we used again (\ref{sommaD}).

\noindent
We are left with an estimate uniform in $\s'$ of the quantity $P(T_\boxplus(\s')>m)$.
Recall that the (zero temperature) dynamics chooses u.a.r. a site and try to update it.
Call $\xi_1$ the time needed to choose for the first time the site 1, then $\xi_2$
the time needed, after the first choose of the site 1, to choose for the first time
the site 2, and so on so forth. Calling $\xi=\sum_{i=1}^L\xi_i$ we have that
$\xi\ge T_\boxplus(\s')$ for all $\s'$ . This is granted by the fact that after the time $\xi_1$
we have definitively that $\s_1=+1$, after the time $\xi_1+\xi_2$
we have definitively that $\s_1=\s_2=+1$ and so on.
Hence we have for all $\s'$
$$P(T_\boxplus(\s')>m)\le P(\xi>m)$$
Being $\xi_i$ a geometrical variable of probability $p=\frac{1}{L}$,
and hence having $E(\xi_i)=L$, $Var(\xi_i)=L^2$ for all $i$,
we have that $\xi$ is the sum of $L$ independent geometric
identical variables, and therefore $E(\xi)=L^2$, $Var(\xi)=L^3$.

\noindent
By Chebyshev inequality
$$
P(\xi>m)=P(\xi-E(\xi)>m-E(\xi))=P(\xi-E(\xi)>m-L^2)\le \frac {L^3}{(m-L^2)^2}
$$
We have then proved that
$$\sup_{\s'}\sum_{m=2L^2+1}^\infty P(T_\boxplus(\s')>m)\le
\sum_{m=2L^2+1}^\infty \frac {L^3}{(m-L^2)^2}\le L$$
which finally gives
\be{sommagrande}
\sum_{\sigma}
 \left| \sum_{\sigma'}\Delta P_{\tau\sigma'}\left(\sum_{m=2L^2+1}^\infty R_m\right)_{\sigma'\sigma}\right|\le 4L
\ee
Combining (\ref{sommagrande}) and (\ref{sommapiccola}) we get (\ref{stima}).

\section{Proof of Theorems \ref{t2} and \ref{t3}}
\label{primord}

Let us denote with $\lambda((k;1),(i;m))$ a sequence of spin flip,
allowed by the zero temperature dynamics, that brings the configuration $(k;1)$
into the configuration $(i;m)$. Since at least one $-$ spin has to be present in all the
steps of the sequence, the latter can be described by partial Dyck words, and the
number of such sequence is given by the elements of the so-called {\it Catalan's triangle}
(see e.g. \cite{Ba}, \cite{W}).

\noindent
We have
$$\pi^{(1)}_{(i;m)}=D_{+, (i;m)}={1\over L}\sum_{k=1}^i\sum_{s=0}^\infty P^{(0)s}_{(k;1),(i;m)}
={1\over L}\sum_{k=1}^i\sum_{s=2i+m-2k-1}^\infty P^{(0)s}_{(k;1),(i;m)}=
$$
$$={1\over L}\sum_{k=1}^i{1\over L^{2i+m-2k-1}}\sum_{\lambda((k;1),(i;m))}\sum_{s'=0}^\infty
{2i+m-2k-1+s'\choose s'}\left(1-{2\over L}\right)^{s'}=
$$
\be{111}=\sum_{k=1}^i{1\over L^{2i+m-2k}}\left({L\over 2}\right)^{2i+m-2k}C_{i+m-k-1,i-k}=
\sum_{k=1}^i\left({1\over 2}\right)^{2i+m-2k}C_{i+m-k-1,i-k}
\ee
where in the second line we defined $s'=s-2i-m+2k+1$, and
in the last line we used the Taylor expansion, convergent for $|\alpha|<1$, of the function
$\left({1\over 1-\alpha}\right)^{N+1}$
$$\left({1\over 1-\alpha}\right)^{N+1}=\sum_{s=0}^\infty
{N+s\choose s}\alpha^s.$$
In equation (\ref{111}) $C_{i+m-k-1,i-k}$ denotes the number appearing in the position
$i+m-k-1,i-k$ of the Catalan's triangle, i.e.
\be{cat}
C_{n,k}={(n+k)!(n-k+1)\over k!(n+1)!}.
\ee
Calling $l=i-k$ we have
\be{cata}\pi^{(1)}_{(i;m)}=\sum_{l=0}^{i-1}\left({1\over 2}\right)^{2l+m}C_{l+m-1,l}\ee
We will now prove the following lemma.
\bl{rw}
For every positive integer $m$ we have
\be{uno}
\sum_{l=0}^\infty \left({1\over 2}\right)^{2l+m}C_{l+m-1,l}=1
\ee
\el

\noindent
{\bf Proof}.
The quantity $\pi^{(1)}_{(i;m)}$ can be written in terms of
a one dimensional Symmetric Random  Walk (SRW),  $S_n=\sum_{i=1}^n X_i$,  with $X_i$ independent
Bernoulli variables $X_i\in \{-1,+1\}$.
Indeed $C_{l+m-1,l}$ is the number of paths of the random walk $\{S_n\}_{n\in\mathbb N}$  such
that $S_1=1, \, S_{2l+m}=m$ and $S_n>0$ for any $n=1,...,2l+m$.
For the duality principle for random walks, we have that $(X_1, X_2,...,X_n)$ has the same distribution
of $(X_n, X_{n-1},...,X_1)$, so that the path $(0, S_1,S_2,...,S_n)$ has the same probability of
the time reversal path $(0,S_n-S_{n-1}, S_n-S_{n-2},...,S_n-0)$.
This implies that
 by denoting with $\t_m$ the first hitting time to $m$ for the
random walk starting at $0$, we have for every positive integer $m$
\be{113} \left({1\over 2}\right)^{2l+m}C_{l+m-1,l}=P(\t_m=2l+m) \ee
so that \be{112}
\pi^{(1)}_{(i;m)}=\sum_{l=0}^{i-1}P(\t_m=2l+m)=P(\t_m< 2i+m). \ee
Formula (\ref{uno}) now immediately  follows from (\ref{112}) since
for the SRW the hitting of any state is finite with probability one.
$\Box$
\vskip.3cm

\noindent
{\bf Remark}. The proof of (\ref{uno}) can also be obtained in a purely combinatorial framework. See for instance
Lemma 18 in reference \cite{ng}.

\noindent
We now prove (\ref{th'2ij}). From (\ref{cata}) and Lemma \ref{rw} we have
\be{new}
\pi^{(1)}_{(i;m)}= 1- \sum_{l=i}^{\infty}\left({1\over 2}\right)^{2l+m}C_{l+m-1,l}
\ee
with
$$
\left({1\over 2}\right)^{2l+m}C_{l+m-1,l}=\left({1\over 2}\right)^{2l+m}{(2l+m)!\over(l+m)!l!}\ {m\over 2l+m}
$$
{ Using upper and lower Stirling's bounds for the factorials \cite{Ro} valid for all $n\ge 1$
$$
\sqrt{2\p n} \left(n\over e\right)^n e^{1\over 12n+1}< n! < \sqrt{2\p n} \left(n\over e\right)^n e^{1\over 12n}
$$}
we have, for any $l\ge 1$ and any $m\ge 1$
$$
\left({1\over 2}\right)^{2l+m}C_{l+m-1,l}\le{  {e^{1\over 12}\over \sqrt{2\p}}}
{\left(1+{m\over 2l}\right)^{2l+m}\over\left(1+{m\over l}\right)^{l+m}}
\ {m\over \sqrt{l(l+m)(2l+m)}}
\le
$$
$$
\le{e^{1\over 12}\over \sqrt{2\p}}
{\left(1+{m\over 2l}\over1+{m\over l}\right)^{m}}{\left(\left(1+{m\over 2l}\right)^2\over1+{m\over l}\right)^{l}}
\ {m\over \sqrt{l(l+m)(2l+m)}}\le $$
$$
\le {e^{1\over 12}\over \sqrt{2\p}} {\left(l+{m\over 2}\over l+{m}\right)^{m}}\left(1+{m\over l}\right)
\ {m\over \sqrt{l(l+m)(2l+m)}}=$$
$$
={{e^{1\over 12}}\over \sqrt{2\p}}\left(1-{m\over 2( l+m)}\right)^m\ {m\over l^{3/2}}\sqrt{l+m\over2l+m}~
$$
$$
\le ~{1\over 2}\,\,e^{- m^2\over 2(m+l)}\; {m\over l^{3/2}}
$$
where in the last line we have used the trivial bound $(1-x)\le e^{-x}$ valid  for all $x\ge 0 $.
Hence for any $l\ge 1$ and any $m\ge 1$ we may roughly bound
\be{stime}
\left({1\over 2}\right)^{2l+m}C_{l+m-1,l}
\le {m\over 2}\,{1\over l^{3/2}}
\ee
A similar  computation gives, for any $l\ge 1$ and any $m\ge 1$,
$$
\left({1\over 2}\right)^{2l+m}C_{l+m-1,l}\ge {e^{-{1\over 6}}\over \sqrt{2\p}}
{\left(1+{m\over 2l}\right)^{2l+m}\over\left(1+{m\over l}\right)^{l+m}}
\ {m\over \sqrt{l(l+m)(2l+m)}}
\ge
$$
$$
\ge{1\over 3}
{\left(1+{m\over 2l}\over1+{m\over l}\right)^{m}}{\left(\left(1+{m\over 2l}\right)^2\over1+{m\over l}\right)^{l}}
\ {m\over \sqrt{l(l+m)(2l+m)}}\ge $$
$$
\ge {1\over 3}{\left(l+{m\over 2}\over l+{m}\right)^{m}}
\ {m\over \sqrt{l(l+m)(2l+m)}}=$$
$$
$$
Therefore we may roughly bound  for any $l\ge 1$ and any $m\ge 1$
\be{stime2}
\left({1\over 2}\right)^{2l+m}C_{l+m-1,l}\ge{2^{-m}\over 3\sqrt{6}}\ {1\over l^{3/2}}
\ee
From inequalities (\ref{stime}) and (\ref{stime2}) the first statement (\ref{th'2ij}) of Theorem \ref{t2} immediately follows.

\noindent
In order to show (\ref{th2ij})
we write
$$\pi^{(1)}_{(i;m)}=\left({1\over 2}\right)^{m}+\sum_{l=1}^{i-1}\left({1\over 2}\right)^{2l+m}{(2l+m)!\over(l+m)!l!}\ {m\over 2l+m}$$
Using now (\ref{stime}) and recalling that $\sum_{n=1}^\infty {1\over n^{3/2}}=\z(3/2)\le 3$,
we get
$$\pi^{(1)}_{(i;m)}\ \le \left({1\over 2}\right)^{m} + {1\over 2}\sum_{l=1}^{i-1}e^{ -m^2\over 2(m+l)} {m\over l^{3/2}}\le \left({1\over 2}\right)^{m}+
{e^{ -m^2\over 2(m+i)} \over 2} \sum_{l=1}^{\infty}{m\over l^{3/2}} \le
$$
\be{piuno}
\le \left({1\over 2}\right)^{m}+
{3}me^{ -m^2\over 2(m+i)}\le (1+ {3}m)e^{ -m^2\over 2(m+i)}\le  {4\,}m\,e^{ -m^2\over 2(m+i)}
\ee
and inserting (\ref{piuno}) inequality into (\ref{new}) we get (\ref{th2ij}).

\noindent
The computation of  $\pi^{(1)}_{(i)}$ is similar,
but it is necessary to choose the time in which the spin in the
site $L$ is flipped to $\s_L=-1$. We have
$$\pi^{(1)}_{(i)}=D_{+, (i)}={1\over L}\sum_{k=1}^i\sum_{m=0}^\infty P^{(0)m}_{(k;1),(i)}=
$$
$$={1\over L}\sum_{k=1}^i\sum_{l=k}^i{1\over L^{L+l-2k-1}}\sum_{\lambda((k;1),(l;L-l))}\sum_{m'=0}^\infty
{L+l-2k-1+m'\choose m'}\left(1-{2\over L}\right)^{m'}\times
$$
$$\times
{1\over L}{1\over L^{i-l}}\sum_{m''=0}^\infty{i-l+m''\choose m''}\left(1-{i\over L}\right)^{m''}=
\sum_{k=1}^i\sum_{l=k}^i\left({1\over 2}\right)^{L+l-2k}C_{L-k-1,l-k}=
$$
$$=\sum_{l=1}^i\sum_{k=1}^l\left({1\over 2}\right)^{L+l-2k}C_{L-k-1,l-k}=
\sum_{l=1}^i\pi^{(1)}_{(l;L-l)}
$$
This ends the proof of Theorem \ref{t2}. $\Box$

\noindent
To prove Theorem \ref{t3}
we first observe that in the chilled regime the Gibbs measure $\p^G(\s)$ is such that
$$
\p^G(\s)={e^{-2J\ell(\s)}\over 1+o(1)}
$$ where
 $o(1)$ denotes any function of $L$ such that
$\lim_{L\to \infty}o(1)= 0$. So if we let 
$$
\widehat \p^G(\s)= e^{-2J\ell(\s)}
$$
we have clearly that
\be{tm3}
\lim_{L\rightarrow\infty}\frac{\pi^{(\le 1)}(m)}{\pi^G(m)} = \lim_{L\rightarrow\infty}\frac{\pi^{(\le 1)}(m)}{\widehat\pi^G(m)}
\ee
%{We then recall that for fixed $m$ and large $i$
%$\pi^{(\le 1)}_{(i;m)}$ tends to $e^{-4J}$, which is the value of
%$\widehat\pi^{G}_{(i;m)}$, but equation (\ref{new}) show that
%\be{posi}
%\pi^{(\le 1)}_{(i;m)}\le \widehat\pi^{G}_{(i;m)}\quad \forall i,m
%\ee
%and when $m$ is of the same order
%of $i$ the value of $\pi^{(\le1)}_{(i;m)}$ is much smaller than
%$e^{-4J}$. This means that $\pi^{(\le 1)}_+\ge \pi^{G}_+$}.
We start computing $\widehat\pi^{G}(m)$.
{
Observe that
\be{equa}
\widehat\pi^{G}(m)= e^{-4J}\sum_{i=1}^L\sum_{m=1}^{L-i}m  + \sum_{m=1}^L m \sum_{k=2}^{L/2}
e^{-4kJ} n(k,m)
\ee
where $n(k,m)$ is the number of configurations with $k$ disjoint intervals of minus spins with a total
number $m$ of minus spins.
Due to the rough estimate $n(k,m)<L^{2k-1}$ we get
%in the chilled regime we can consider the first order
%approximation also for the Gibbs measure obtaining
\be{equa1}
\widehat\pi^{G}(m)\le \left[{e^{-4J}\over 6}\left({L^3}-{L}\right)  + L^3e^{-4J}o(1)\right]\le\frac{L^3e^{-4J}}{6}( 1+o(1))
\ee
%\be{equa1}
%\pi^{G}(m)=
%e^{-4J}\sum_{i=1}^L\sum_{m=1}^{L-i}m+L^3e^{-4J}o(1)={e^{-4J}\over 6}\left({L^3}-{L}\right)+L^3e^{-4J}O(L^{-4\g})\le
%\frac{L^3e^{-4J}}{6}\Big( 1+O(L^{-4\g})\Big)
%\ee}
%Call now $\D{\bar m}$ the quantity
%$$
%\D{\bar m}=\sum_{i=1}^L\sum_{m=1}^{L-i}(\pi^{G}_{(i;m)}- \pi^{(\le 1)}_{(i;m)})m
%$$
%We want to show that it can be found a constant $c$ such that
%$$
%\D{\bar m}\ge c\ e^{-4J}{L^3}
%$$
We next estimate the difference $\widehat\pi^{G}(m)-\pi^{(\le 1)}(m)$. Observe  that  by (\ref{equa})
$$
\widehat\pi^{G}(m)\ge e^{-4J}\sum_{i=1}^L\sum_{m=1}^{L-i}m
$$
and  that by (\ref{terminek}) and (\ref{th2ij})
$$
\pi^{(\le 1)}(m)= \pi^{(\le 1)}_{(i;m)}= e^{-4J}\pi^{(1)}_{(i;m)}
$$
so we have
$$
\widehat\pi^{G}(m)-\pi^{(\le 1)}(m)\ge e^{-4J}\sum_{i=1}^{L}\sum_{m=1}^{L-i}m(1- \pi^{( 1)}_{(i;m)})
$$
Then note that, due to  (\ref{new}) we have that $1- \pi^{( 1)}_{(i;m)}>0$, so
we are allowed to restricted the sums over $i,m$  above to a subset in which $i\le m$. Recalling also 
bound (\ref{th2ij}) we get 
$$
\widehat\pi^{G}(m)-\pi^{(\le 1)}(m)\ge  e^{-4J}
\sum_{i=1}^{L/2}\sum_{m=i}^{L-i}m(1- \pi^{( 1)}_{(i;m)})\ge e^{-4J}\sum_{i=1}^{L/2}\sum_{m=i}^{L-i}\left(m-4 \;e^{-\frac{m^2}{2(i+m)}}~m^2\right)
$$
\be{fina}
~~~~~~~~\ge e^{-4J}\sum_{i=1}^{L/2}\sum_{m=i}^{L-i}\left(m-4 \;e^{-\frac{m}{4}}~m^2\right)\ge \frac{e^{-4J}L^3}{8}(1+o(1))
\ee
%\widehat\pi^{G}(m)-\pi^{(\le 1)}(m)\ge\sum_{i=1}^{L}\sum_{m=1}^{L-i}(\widehat\pi^{G}_{(i;m)}- \pi^{(\le 1)}_{(i;m)})m\ge  \sum_{i=1}^{L/2}\sum_{m=i}^{L-i}(\widehat\pi^{G}_{(i;m)}- \pi^{(\le 1)}_{(i;m)})m\ge$$
%$$
%\ge e^{-4J}\sum_{i=1}^{L/2}\sum_{m=i}^{L-i}\left(m-4 \;e^{-\frac{m^2}{2(i+m)}}~m^2\right) \ge e^{-4J}\sum_{i=1}^{L/2}\sum_{m=i}^{L-i}\left(m-4 \;e^{-\frac{m}{4}}~m^2\right)
%$$
%{\red Now note that
%$$
%\sum_{i=1}^{L/2}\sum_{m=i}^{L-i} m= {L^3\over 8}
%$$
%Moreover, for any $m\ge 1$
%$$
% m^2 e^{-\frac{m}{4}}\le{256\over e^2}e^{ -\frac{m}{8}}\le 35 e^{ -\frac{m}{8}}
% $$
% So that
% $$
% \sum_{i=1}^{L/2}\sum_{m=i}^{L-i}m^2 e^{-\frac{m}{4}}\le 35 \sum_{i=1}^{L/2}\sum_{m=i}^{L-i}
% e^{ -\frac{m}{8}}\le  35\sum_{i=1}^{\infty}\sum_{m=i}^{\infty} e^{ -\frac{m}{8}}\le 35 \sum_{i=1}^{\infty} 9e^{-{i\over 8}}\le 35\cdot 9\cdot 8=2520
% $$}
%Therefore
%\noindent
% It is simple to conclude that there is a constant $C'$ such that $\widehat\pi^{G}(m)-\pi^{(\le 1)}(m)
%\ge e^{-4J}\left(\frac{L^3}{8}- C'\right)$
Hence, from inequalities (\ref{equa1})  and (\ref{fina}) we get
$$
{\widehat\pi^{G}(m)-\pi^{(\le 1)}(m)\over \widehat\pi^{G}(m)}\ge {3\over 4}\left(1+o(1)\right)
$$
whence
$$
\lim_{L\to \infty} {\pi^{(\le 1)}(m)\over  \widehat\pi^{G}(m)} \le {1\over 4}
$$
and from (\ref{tm3})
Theorem \ref{t3} immediately follows.
 $\Box$

\section{Conclusions}
\label{conc}
In this paper we have considered an example of a
single spin flip irreversible dynamics
for a system very simple, but yet quite difficult to study in presence of boundary
conditions.
With explicit estimates we have shown that, expanding in series
the stationary measure around the zero temperature, it is possible to control
for very low temperature
the convergence of the expansion and to compute, up to the first order,
the stationary probability distribution.
The latter has non trivial features: it has an explicit dependence both
on the relative distance and on the position of the changes of sign in the
state.
Moreover the memory of the boundary conditions has a very slow decay,
and crucial macroscopic effects.

\noindent
There are several questions opened by this result.
The generalization of this computations to PCA dynamics, like the one
presented in \cite{dss1} and \cite {dss2}, should be straightforward.
It should be possible also, with some extra effort, to understand the features of the
higher terms of the expansion, and it would be very interesting to generalize this
technique to higher dimensions. All these questions will be the subject of further investigations.

\vskip.5truecm
\noindent
{\bf Acknowledgments:}
We thank Pietro Caputo for equation (\ref{113}) and Francesco Pappalardi for
useful discussions on the combinatorial properties of Catalan's triagles, and
related references. We also thank A.De Sole, D.Gabrielli, G.Gallavotti, G.Jona-Lasinio
for useful discussions.
A.P.  has been partially supported by
Conselho Nacional de Desenvolvimento Cient\'\i fico e Tecnol\'ogico
(CNPq), Funda\c c\~ao de Amparo \`a  Pesquisa do Estado de Minas Gerais (FAPEMIG - Programa de Pesquisador Mineiro)
and  by the Simons Foundation
and  the Mathematisches Forschungsinstitut Oberwolfach. His stay in Rome during part of this work has been founded by
the Grant ``Visiting Professors'' of the Universita' di Roma Tor Vergata.
B.S. and E.S. thank
the support of the A*MIDEX project (n. ANR-11-IDEX-0001-02) funded by the  ``Investissements d'Avenir" French Government program, managed by the French National Research Agency (ANR).
B.S. has been supported by PRIN 2012, Problemi matematici in teoria cinetica ed
applicazioni. E.S. has been supported  by the PRIN 20155PAWZB ``Large Scale Random Structures".

\end{document}